# CAN DAMAGE OBSERVATIONS FROM THE 2019 LE TEIL (FRANCE) EARTHQUAKE CHALLENGE ESHM20 AND ESRM20?

K. Trevlopoulos[1], P. Gehl[2] & C. Negulescu[3]

[1] BRGM, F-45060 Orléans, France, k.trevlopoulos@brgm.fr
[2] BRGM, F-45060 Orléans, France, p.gehl@brgm.fr
[3] BRGM, F-45060 Orléans, France, c.negulescu@brgm.fr

**Abstract**: *Probabilistic seismic hazard and risk models are essential to improving our awareness of seismic risk, to its management, and to increasing our resilience against earthquake disasters. These models consist of a series of components, which may be tested and validated individually, however testing and validating these types of models as a whole is challenging due to the lack of recognised procedures. Estimations made with other models, as well as observations of ground shaking and damages in past earthquakes lend themselves to testing the components for ground motion modelling and for the severity of damage to buildings. Here, we are using observations of damages caused by the Le Teil 2019 earthquake, third-party estimations of macroseismic intensity for this seismic event, and ShakeMap analyses in order to make comparisons with estimations made with scenario simulations using model components developed in the context of the 2020 Euro-Mediterranean Seismic Hazard Model and the European Seismic Risk Model. The comparisons concern the estimated ground motion intensity measures, the macroseismic intensity, the number of damaged buildings, and the probabilities of the damage grade. The divergences of the estimations from the observations, which are observed in some of comparisons, are attributed to factors external to the models, such as the location of the hypocentre.*

## 1 Introduction

Earthquakes are among the disasters with most severe consequences, which include loss of human life, disruption of critical infrastructures, insured and uninsured losses, indirect economic losses, as well as socio-technical impacts in multi-risk safety contexts. Assessments based on probabilistic seismic hazard and risk analysis (PSHA, PSRA) are key elements of efforts to improve awareness of seismic risk, response, and resilience to earthquakes. As far as seismic hazard and risk in Europe is concerned, the 2020 European Seismic Hazard and Risk Models (ESHM20, ESRM20 - Crowley et al., 2021a; Danciu et al., 2021) are the state of the art models, which were created by the European Facilities for Earthquake Hazard and Risk consortium. The predictive accuracy of the multi-component ESHM20 and ESRM20 models, as that of all seismic hazard and risk models, and as that of all statistical and probabilistic models, needs to be tested, despite the fact that the individual components consisting them have already undergone testing.

In the nuclear industry, testing and evaluation of PSHA models and their components have been formalized in the form of Senior Seismic Hazard Analysis Committee (SSHAC) Hazard Studies (Ake et al., 2018). SSHAC projects aim to produce "technically defensible" distributions and probabilities of exceedance of ground motion intensity measures. Bommer et al. (2013) tested ground motion models and their logic tree by comparing their implementations by three independent teams of modellers. As far as the evaluation of PSHA logic trees is concerned, Marzocchi et al. (2015) argue that the hazard should be considered to be an ensemble of models, which do not need to be mutually exclusive and collectively exhaustive. Rood et al. (2020) used observations of geomechanical failures, i.e., rock toppling, to estimate upper limits of ground motion intensity measures and constrain hazard estimations for long return periods. Their procedure always leads to a reduction of the seismic hazard estimation, which depends on the model for the seismic fragility, i.e., the model estimating the probability of geomechanical failure conditioned on a ground motion intensity measure. Moreover, they proposed a procedure for dropping branches of the PSHA logic tree and reweighting the remaining. Gerstenberger et al. (2020) note that tests of national or regional hazard models are only meaningful at the level of the site, and that resorting to conversions of macroseismic intensity to ground motion intensity, when ground motion records are lacking, may introduce errors. Nevertheless, Mak and Schorlemmer (2016) did use such a conversion after testing the conversion equation itself.

In this study, to make comparisons using components of the ESHM20 and the ESRM20, we use observations of damage in buildings in the municipality of Le Teil, France, caused by the 2019 Le Teil earthquake. First, we generate samples for a set of ground motion intensity measures (IMs) given by scenario simulations, based on the ESHM20 for different hypocentres and focal mechanisms reported by different sources. The distributions of the samples are compared to distributions given by ShakeMap analyses (Wald et al., 2022), in order to select the most compatible scenario simulation. We convert the IMs to macroseismic intensities using different ground-motion intensity conversion equations (GMICEs). A third-party macroseismic intensity estimation for the municipality of Le Teil is then used to select the most plausible scenario simulation. Subsequently, we consider alternative exposure models, and $V_{S30}$ models, and we estimate the probabilities of the damage states of the buildings, which we compare to the corresponding probabilities based on damage observations and expert judgement.

## 2 Seismological and damage data

### 2.1 Seismic hazard and risk, and information for 2019 Le Teil earthquake

The municipality of Le Teil is located in southeastern metropolitan France, a region that corresponds to low and moderate risk categories, according to the French Seismic Zonation. For Le Teil in particular, the ESHM20 estimates a mean Peak Ground Acceleration (PGA) of 0.04 g with a 0.21 % probability of exceedance in 1 year (475 years mean return period) on rock site conditions ($V_{S30}$ = 800 m/s).

The Le Teil earthquake took place on the 11th of November 2019, and its epicentre is located at 44.518° N 4.671° E (Ritz et al., 2020) in close proximity to the municipality of Le Teil and the town of Montélimar in the Lower Rhône valley in France. A private power plant accelerometer, located 15 km north-northeast of the epicentre, recorded PGA of 0.045 g (Schlupp et al., 2022), as the closest seismic station to the earthquake. Three stations of the French seismological and geodetic network (Résif / EPOS-FR) at 24-44 km from the epicentre recorded PGAs in the range of 0.004-0.007 g. These four stations are at such a distance from the epicentre and the municipality of Le Teil, so that they cannot accurately constrain the predicted IMs. (Causse et al., 2021) used numerical modelling, including physics-based rupture modelling and modelling of near-fault wave propagation, and estimated near-fault PGAs with a 68 % confidence interval of 0.3-1.9 g. They argued that their estimations are compatible with displacements of rigid block objects such as rocks and ledger stones. Moreover, they suggested that existing ground motion models may not be useful in the case of earthquakes such as this one, with a rarely observed shallow hypocentral depth, and with rupture parameters such as stress drop that are usually associated with earthquakes not only at larger depths, but of larger magnitudes too.

Schlupp et al. (2022) reported an EMS98 macroseismic intensity of 7-8 for the municipality of Le Teil. This conclusion was the product of expert judgement considering the EMS98 definitions of the intensity degrees and damage grades, the field observations from the Macroseismic Response Group, and the EMS98 vulnerability classes of the buildings based on land registration data. Based on this procedure, Schlupp et al. (2022) determined 765 macroseismic intensities covering the area affected by the earthquake. The isoseist line of the map by Schlupp et al. (2022) for intensity VII includes the built area of the Le Teil: given the limited spatial extent of this area, there is practically no spatial variation of the macroseismic intensity within this isoseist line, and the maximum is at the Le Teil (7.5).

### 2.2 Damage observations data set

We produced the data set used here by processing post-seismic inspection forms, and by completing and editing an existing data set (Perez, 2020). The inspection forms were filled in by the French Association of Earthquake Engineering (AFPS) during on-site inspections (Taillefer et al., 2021), which took place from the 3rd to the 5th of February 2020. The produced data set contains 327 entries with information about the coordinates of each inspected building, the number of storeys, the date of construction, the degree of damage for the entirety of each inspected building as well as individual damage degree tags for structural and non-structural components. The degree of damage in the observations is on a three-level scale, i.e. green-yellow-red, which we converted to EMS98 damage grades.

For the conversion of the damage observations data in the forms, we used the rules in Table 2 1. We defined these rules based on expert judgement, and they are based on the observed structural and non-structural damage, which are the criteria for classification according to the EMS98 damage scale (Grünthal, 1998). Therefore, the data in the forms, that we used, are the entries in the fields for the structural elements bearing vertical and horizontal loads (which were considered separately), and for the non-structural elements as well. The rest of the fields on the forms are related to procedures for life safety, e.g. evacuation, and they were not required for classifying damage according to the EMS98. In this way, we used the raw information from the inspection forms to classify buildings according to structural damage and not whether a building was usable or not. The results of this reclassification, which involves the distribution of EMS98 damage levels in the green, yellow and red labels, are presented in Table 2-2 for the entire dataset independent of building typology.

*Table 2-1 Proposed classification of the observed damage in the EMS-98 damage grades as a function of the colour tags assigned by the inspectors.*

| Type of elements | Colour tag: G (green), Y (yellow), R (red) | | | | | | | | | | | | | | | |
|---|---|---|---|---|---|---|---|---|---|---|---|---|---|---|---|---|
| Vertical loads-bearing structural elements | R | | | Y | Y | Y | Y | G | G | Y | Y | G | G | G | G | G |
| Horizontal loads-bearing structural elements | | R | | Y | Y | Y | Y | Y | Y | G | G | G | G | G | G | G |
| Internal non-structural elements | | | R | R | Y | R | Y | R | Y | R | Y | R | Y | Y | G | G |
| External non-structural elements | | | | R | R | R | Y | Y | R | R | R | R | Y | R | Y | Y | G |
| EMS-98 damage grade | 5 | 4 | 2 | 2 | 4 | 4 | 3 | 3 | 3 | 3 | 4 | 3 | 2 | 2 | 2 | 2 | 1 |

*Table 2-2 Percentage of buildings in each damage grade as a function of the building's tag for the entire dataset*

| Building tag | Damage grade | Count | Percentage (%) |
|---|---|---|---|
| Green | 1 | 91 | 61 |
| Green | 2 | 22 | 15 |
| Green | 3 | 35 | 24 |
| Yellow | 3 | 95 | 90 |
| Yellow | 4 | 8 | 8 |
| Yellow | 5 | 2 | 2 |
| Red | 4 | 47 | 64 |
| Red | 5 | 27 | 36 |

## 3 Comparisons using the models for seismic hazard and risk

### 3.1 Comparisons based on the intensity of the seismic ground motion

Here we compare the macroseismic intensity reported by Schlupp et al. (2022) to that resulting from ShakeMap analyses, and scenario analyses. The scenario analyses are conducted for five different rupture models using the OpenQuake Engine (Pagani et al., 2014; Silva et al., 2014) and the ground-motion prediction equation (GMPE) "KothaEtAl2020Site", a version of the GMPE by Kotha et al. (2020) with a polynomial site amplification as a function of the $V_{S30}$, which is available in the OpenQuake Engine. The geometries of the ruptures in the ShakemMap analyses as well as in the scenario analyses are all modelled as "Simple Faults" of flat square geometry, each defined by the set of parameters in Table 3-1. The scenarios are named after the source of the data for the magnitude and the hypocentre location, i.e., "CEA" (CEA/LDG, 2011; Duverger et al., 2021), "EMSC" (EMSC, 2011), "RENASS" (BCSF-RENASS, 2011), "Ritz et al." (Ritz et al., 2020) and "USGS" (USGS, 2011). The strike, dip, and rake angles of the focal mechanism solutions reported by "CEA" and "Ritz" are arbitrarily assigned to the scenarios "EMSC" and "RENASS", respectively. The surface of the rupture is estimated using the Wells and Coppersmith (1994) scaling law, and the coordinates of the points defining the rupture geometry are calculated in order to be used in the OpenQuake Engine simulations and in the conversion of ground motion IMs to macroseismic intensity. To calculate the coordinates of the corners of the rupture geometry, we assume that its centre of gravity is located at the hypocentre. This assumption leads in some cases to an upper rupture edge above ground surface. This is amended by translating the rupture geometry on its plane so that its upper edge coincides with the fault trace on ground surface. The depths of the upper and lower edges of the rupture geometry are used to define in the Simple Fault model the upper and lower seismogenic depths, respectively. The coordinates of the ends of the trace of the fault on the ground surface required by the Simple Fault model are calculated by projecting the rupture geometry on the ground surface in the direction of the dip. Moreover, a maximum rupture mesh spacing of 0.5 km is used, which leads to a 6 by 6 grid in all scenario analyses, which we consider sufficient.

*Table 3-1 Rupture assumptions used in the five source models*

| Scenario name | $M_W$ | Hypocentre longitude (°E) | Hypocentre latitude (°N) | Hypocentre depth (km) | Strike (°) | Dip (°) | Rake (°) |
|---|---|---|---|---|---|---|---|
| "CEA" | 4.9 | 4.65 | 44.53 | 2.0 | 47 | 65 | 93 |
| "EMSC" | 4.9 | 4.62 | 44.57 | 10.0 | 47 | 65 | 93 |
| "RENASS" | 4.8 | 4.64 | 44.53 | 2.0 | 50 | 45 | 89 |
| "Ritz et al." | 4.9 | 4.671 | 44.518 | 1.0 | 50 | 45 | 89 |
| "USGS" | 4.84 | 4.638 | 44.612 | 11.5 | 53 | 57 | 99 |

To account for the uncertainty in the intensity of the ground motion, 1000 ground motion fields are generated, i.e. samples of IMs at a series of geographic points, which include the centroids of the exposure model. The ground motion fields are generated for the IMs peak ground acceleration (PGA), spectral pseudo-acceleration at 0.3, 0.6, 1.0 and 3.0 s. Furthermore, the spatial correlation of the IMs is taken into account in the generation of the IM samples by using the Jayaram and Baker (2009) model in the OpenQuake Engine, assuming no clustering of the $V_{S30}$ values in the study area.

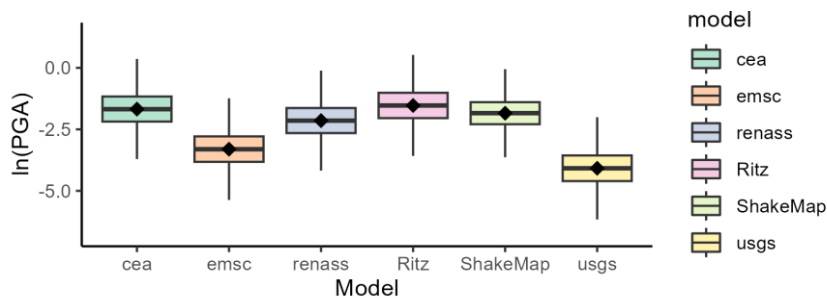

*Figure 1. Boxplots for the generated samples for the considered ground motion intensity measures at all exposure centroids based on the scenario simulations (the edges of the box are located at the first and third quartile, respectively, the line at the middle of the box is located at the median, the point marker is located at the mean of the sample, the whiskers extend up to 1.5 times the distance between the first and third quartile approximating the 95 % confidence interval).*

Figure 1 shows box plots for the samples generated for the PGA aggregated over all exposure centroids. If we consider only the boxplots corresponding to the five scenarios in Table 4 ("CEA", "EMSC", "RENASS", "Ritz et al.", "USGS"), the dispersions of the samples are equivalent, as expected due to the use of the same GMPE. However, the differences with respect to the means of these five IM samples has to be attributed to the differences between the epicentre locations, the depth of the hypocentre, and the focal solution, because these are the parameters affecting the distance between the exposure centroids and the geometry of the rupture. Moreover, the means for the scenarios "EMSC" and "USGS" are consistently the lowest. We attribute this primarily to the hypocentral depths in these two scenarios (10.0 and 11.5 km), which are significantly larger those in the other 3 scenarios, leading to distances from the rupture between 10.0 and 25.0 km, when the corresponding distances in the other 3 scenarios are less than 5.0 km. As far as the boxplot for the samples based on the ShakeMap analysis is concerned, the boxplot whiskers are relatively shorter than those for the 5 scenarios, signifying smaller dispersions of the IM logarithms. This difference should primarily originate from the differences between the GMPEs in the ShakeMap configuration and in the scenario simulations.

### 3.2 Comparisons based on the macroseismic intensity

The generated IM samples are subsequently converted to macroseismic intensities using Ground Motion to Intensity Conversion Equations (GMICEs) and are subsequently compared with the macroseismic intensity reported by Schlupp et al. (2022). The aim of this comparison is to identify the scenarios leading to macroseismic intensities closest to the reported. To this end, we use two GMICEs, which we consider compatible with the study area. These are the GMICEs by Faenza and Michelini (2010) ("FM2010") and by Caprio et al. (2015) ("CA2015").

Figure 2 shows the boxplots for the MCS and the INT, respectively, which resulted from the conversion of the IM samples. Despite the fact that the MMI and MCS have differences, we adopt here the guidelines by Musson et al. (2010), which take the two scales as equivalent (to each other and to the EMS-98 scale) up to intensity 10. We make this assumption to distinguish the effects of the employed GMICEs on the distributions of the generated samples of macroseismic intensities in Figure 2 from the differences due to the underlying hazard model components.

In order to assess the usefulness of the distribution for each scenario in Figure 2, we are using the 7.5 EMS-98 intensity estimated by Schlupp et al. (2022) for the municipality of Le Teil. The MCS distributions resulting from the FM2010 model, whose median is closer to the 7.5 observation-based estimation, are those for the CEA, RENASS, and Ritz et al. scenarios, and USGS ShakeMap analyses. As far as the application of the CA2015 model (not shown here) is concerned, it leads to macroseismic intensity distributions with larger dispersions and lower medians with respect to the FM2010 (Figure 2) in the cases considered. In the cases examined here, the distributions whose median closest to the 7.5 observation-based estimation, are those for the scenarios CEA, RENASS, and Ritz et al., and the distributions from the ShakeMap analyses.

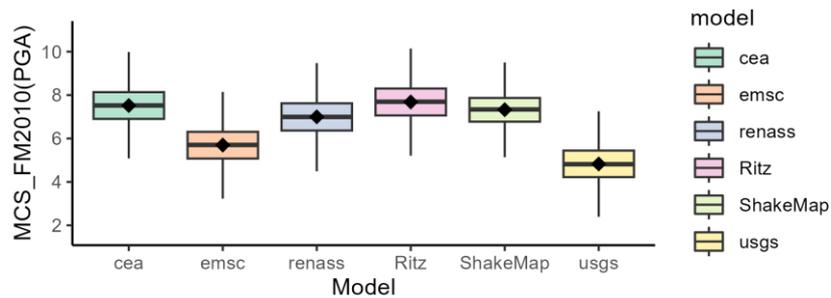

*Figure 2. Boxplots for a) the Mercalli-Cancani-Sieberg (MCS) macroseismic intensity as a function of the PGA given by the ground motion-to-intensity conversion equation by Faenza and Michelini [2010] ("FM2010") (the edges of the box are located at the first and third quartile, respectively, the line at the middle of the box is located at the median, the point marker is located at the mean of the sample, the whiskers extend up to 1.5 times the distance between the first and third quartile approximating the 95 % confidence interval).*

### 3.3 Comparisons based on the probability of damage

*Estimation of damage using different risk analysis tools with equivalent exposures*

For our first comparison with respect to the number of damages in the municipality of Le Teil, we compare the estimated damages using the seismic risk analysis tool Armagedom (Sedan et al. 2013) on the VISIRISKS platform (Negulescu et al. 2023) with an estimation made with a scenario analysis with the OpenQuake Engine using the ESHM20 ground motion modelling logic tree, and elements of the ESRM20. As far as Armagedom is concerned, it implements the semi-empirical macroseismic method by Lagomarsino et al. (2006). For this comparison, we use exposure and fragility models, which we consider equivalent so as to limit the effect of these two factors on the differences between the two estimations. The exposure model used in the analysis with the OpenQuake Engine was created based on the exposure model in Armagedom, which includes 2778 buildings of 12 building classes located at 9 centroids across the municipality of Le Teil. For each building class in the exposure model in Armagedom, there is a set of probabilities with respect to how the buildings in that class are distributed to one of the EMS-98 vulnerability classes. Based on this exposure model, we calculated the number of buildings in each vulnerability class, and we selected a fragility model (Table 3-3), which resulted in a simplified exposure model, which approximates the exposure model in Armagedom and which is used in the scenario analysis with the OpenQuake Engine. The estimated number of buildings in each structural damage grade (No damage / Slight / Moderate / Heavy / Very Heavy) based on the two analyses is given in Figure 3. The percentage of buildings with Heavy and Very Heavy damage is 0.9 % and 1.2 % based on the analysis with Armagedom and the analysis with the OpenQuake Engine, respectively. Although these results are lower than the observed Heavy and Very Heavy damage in Le Teil, they show that the components of the two approaches lead to similar results in this case.

*Table 3-3 ESRM20 fragility models selected as corresponding to the buildings in the exposure model used in the simulation using Armagedom based on their EMS-98 vulnerability class*

| EMS-98 Vulnerability Class | Selected ESRM20 Fragility Model |
|---|---|
| A | MUR-STDRE_LWAL-DNO_H3 |
| B | MUR-STDRE_LWAL-DNO_H1 |
| C | MCF_LWAL-DUL_H2 |
| D | MR_LWAL-DUH_H2 |
| E | MR_LWAL-DUH_H1 |
| F | CR_LDUAL-DUH_H2 |

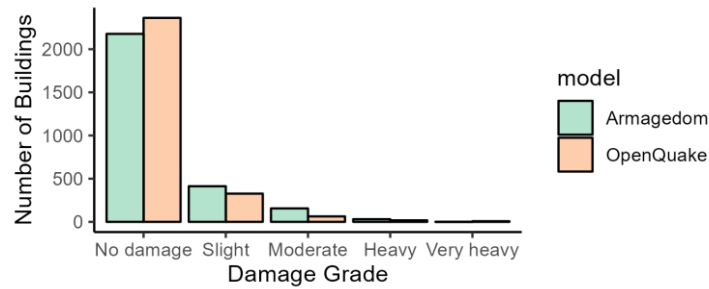

*Figure 3. Estimated number of buildings in each damage grade based on two scenario analyses using the Armagedom risk analysis tool on the VISIRISKS platform (Armagedom) and a model with equivalent exposure and fragility models (selected from the ESRM20) combined with elements of the ESHM20*

*Damage based on observations*

For the comparisons related to vulnerability and risk modelling, we employ simulations using elements of the ESHM20 and the ESRM20 in combination with the dataset produced based on the damage observations. In the simulations, the fragility model consists of fragility curves from the ESRM20, which we selected according to the information in the damage dataset. Initially, we selected a GED4ALL building class based on the building materials and the number of storeys). Moreover, we assigned an EMS98 vulnerability class according to the building material and year of construction. Then, we selected the ESRM20 classes in Table 3-4, as well as the corresponding fragility curves for the simulations, based on the GEDALL class and the EMS98 vulnerability class.

*Table 3-4 Assigned GED4ALL, ESM98 vulnerability, and ESRM20 building classes for the buildings in the AFPS damage observations dataset. The fragility curves in ESRM for the selected classes are function of the listed intensity measure types (IMT)*

| GED4ALL class | EMS98 vuln. class | ERSM20 class | IMT | Number of buildings |
|---|---|---|---|---|
| MUR/LWAL+DNO/HAPP:2 | A | MUR-STDRE_LWAL-DNO_H2 | $S_a(0.3s)$ | 124 |
| MUR/LWAL+DNO/HAPP:2 | B-D | MCF_LWAL-DUL_H2 | PGA | 20 |
| MUR/LWAL+DNO/HAPP:4 | A | MUR-STDRE_LWAL-DNO_H3 | $S_a(0.6s)$ | 122 |
| MUR/LWAL+DNO/HAPP:4 | B,D | MCF_LWAL-DUL_H3 | $S_a(0.3s)$ | 6 |
| CR/LFINF/HAPP:2 | C | CR_LFINF-CDL-10_H2 | $S_a(0.6s)$ | 23 |
| CR/LFINF/HAPP:2 | E-D | CR_LFINF-CDM-0_H1 | $S_a(0.3s)$ | 2 |
| CR/LFINF/HAPP:4 | C | CR_LFINF-CDL-15_H4 | $S_a(1.0s)$ | 29 |
| CR/LFINF/HAPP:4 | E | CR_LFINF-CDM-10_H1 | $S_a(0.3s)$ | 1 |

*Estimated damage based on a "building-by-building" exposure model*

Subsequently, we perform "scenario damage" simulations using the OpenQuake Engine, in which the exposure model assumes that each building is located at the coordinates in the damage dataset in a "building-by-building" sense, and includes 327 buildings of the ESRM20 classes in Table 3-4 (labelled "Sim – brgm $V_{S30}$" in Figure 4). These are the buildings for which the information in the dataset is sufficient for determining the building class and damage grade. The fragility model is defined using the fragility curves from the ESRM20 for the building classes in Table 3-4, while the rupture is modelled according to the "Ritz et al." scenario (Table 3-1). The same type of simulation is performed using ground motion fields generated using parameters of the ground motion intensity measures, which are computed with a ShakeMap analysis, following the procedure described in Section 3.1 (labelled "SM – brgm $V_{S30}$" in Figure 4). In Figure 4, which gives the probability of the damage grades based on the simulations, we see that the "SM – brgm $V_{S30}$" simulation leads to lower probabilities for the damage grades 3-5 than the "Sim. – brgm $V_{S30}$" simulation. The main drivers of the probabilities of the damage grades are the buildings in the classes MUR-STDRE_LWAL-DNO_H2 and MUR-STDRE_LWAL-DNO_H3, which include 38 % and 37 %, respectively, of the total number of buildings in the model. These two classes are also the most vulnerable among the classes in the model, as indicated by the fact that they were classified in the EMS98 vulnerability class A. The fragility curves of these two building classes are functions of Sa(0.3s) and Sa(0.6s), respectively. Based on the results in Figure 1, we consider that the Sa(0.3s) is on average higher in the "Sim – brgm $V_{S30}$" simulation than in "SM – brgm $V_{S30}$", and that there are no significant differences between the two with respect to the Sa(0.6s). This is the factor to which we attribute the differences in the probabilities of the damage grades based on the simulations "Sim. – brgm $V_{S30}$" and "SM – brgm $V_{S30}$".

The effect of the $V_{S30}$ mapping on the estimated probabilities of the damage grade is investigated by using two different site models. The first site model ("brgm $V_{S30}$") is configured based on the $V_{S30}$ values extracted from BRGM's $V_{S30}$ database (Weatherill et al., 2023), which correspond to the coordinates of the buildings in the exposure model. This is the site model used in the simulations "Sim. – brgm $V_{S30}$" and "SM – brgm $V_{S30}$". For the second site model, which was used in the simulation "Sim. – ESHM20 $V_{S30}$", the $V_{S30}$ values are obtained by using the "exposure to site tool" in the ESRM20, in which the "point" workflow is applied, which returns the $V_{S30}$ values at the exact coordinates of the buildings. The simulation "Sim. – ESHM20 $V_{S30}$" leads to the lowest probabilities for the damage grades 3-5 amongst all computations in Figure 4. In this simulation, 68 % of the buildings are located on sites with $V_{S30} \geq 800$ m·s$^{-1}$, while in "Sim. – brgm $V_{S30}$" 72 % of the buildings are on sites with $V_{S30} \leq 360$ m·s$^{-1}$, which is expected to lead to higher ground motion intensities due to site amplification.

Figure 4 also includes the probabilities of the damage grades based on our conversion of the damage observations. For damage grades 4 and 5, there are significant differences between the probabilities based on this approach and the corresponding probabilities based on the scenario simulations and the Shakemap analysis, however, they are not as important as the differences in the case of the damage grades 2 and 3. We presume that the rule that we used for the translation of the damage observations to damage grades (Table 2-1) is the source of these discrepancies.

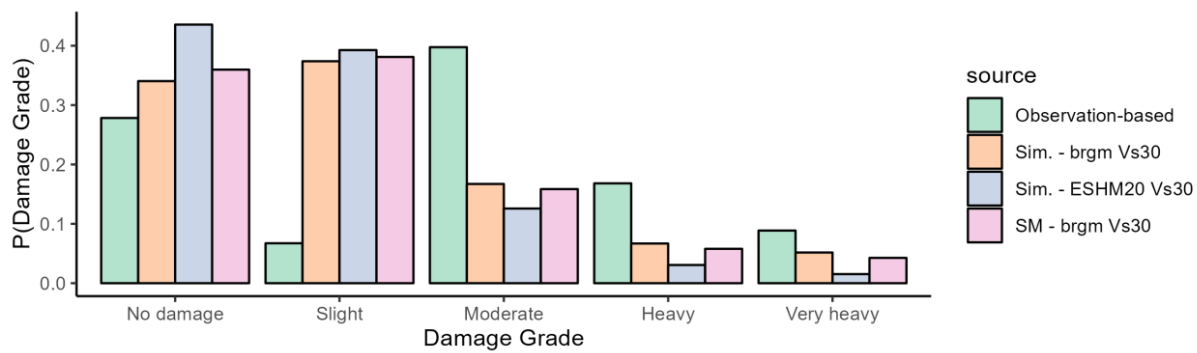

*Figure 4. Number of buildings in the damage grades Moderate to Very heavy according to the model BRGM/CCR, the model based on the ESRM20 ("model"), and based on the field observations ("observations").*

*Estimated damage based on aggregated exposure models*

In addition to the simulations, where the exposure model included 327 inspected buildings at their precise coordinates, we perform a series of "scenario damage" simulations with two aggregated exposure models including the total number of buildings in the municipality of Le Teil. The first exposure model ("ESRM20 exp."), which is based on the ESRM20 exposure (Crowley et al., 2019, 2020, 2021b), includes a single centroid and a total of 1679 buildings. This exposure model results by simplifying the ESRM20 exposure model by fusing similar building types with a small portion of the overall number of buildings in the original ESRM20 exposure (Table A1) into 7 building classes (Table A2). The second exposure model ("brgm exp.") is based on national statistical data, and includes 9 centroids with 2778 buildings. In this exposure model, the buildings are categorized in 12 ESRM20 classes, which we selected based on the exposure model in Sedan et al. (2013).

As far as site models are concerned, four different models are used, two for each exposure model. The site model "ESHM $V_{S30}$", which is used in combination with the exposure model "ESRM20 exp.", takes into account the value of the $V_{S30}$ (834 $m·s^{-1}$) at the coordinates of the exposure centroid, which results by using the "exposure" workflow in the "exposure to site tool" in the ESRM20. Based on this workflow, the value of the $V_{S30}$ is calculated at the coordinates of the exposure centroid by averaging over the polygon of the municipality of Le Teil (République Française, 2022). For the second site model ("brgm $V_{S30}$") used in combination with the "ESRM20 exp." model, we use the $V_{S30}$ value (270 $m·s^{-1}$) in BRGM's $V_{S30}$ database for the coordinates of the centroid.

The $V_{S30}$ values in the two site models for the coordinates of the centroids in the "brgm exp." exposure model are compared in Table 3-5. The values of the $V_{S30}$ in the model "ESHM $V_{S30}$" were calculated using the "point" workflow in the "exposure to site tool". The "brgm $V_{S30}$" model includes $V_{S30}$ values corresponding to soft soils, while the lowest $V_{S30}$ values in the "ESHM $V_{S30}$" model are typical of hard soil sites.

*Table 3-5 $V_{S30}$ at the exposure centroids in the site models "ESHM $V_{S30}$" and "brgm $V_{S30}$"*

| Centroid | Longitude | Latitude | ESM20 $V_{S30}$ ($m·s^{-1}$) | brgm $V_{S30}$ ($m·s^{-1}$) |
|---|---|---|---|---|
| 0 | 4.6835 | 44.5546 | 807 | 800 |
| 1 | 4.6804 | 44.5453 | 831 | 270 |
| 2 | 4.6846 | 44.5414 | 730 | 270 |
| 3 | 4.6498 | 44.5405 | 726 | 800 |
| 4 | 4.6713 | 44.5347 | 831 | 800 |
| 5 | 4.6909 | 44.5500 | 699 | 270 |
| 6 | 4.6699 | 44.5442 | 830 | 800 |
| 7 | 4.6692 | 44.5547 | 840 | 580 |
| 8 | 4.6953 | 44.5315 | 270 | 644 |

The mean probabilities of the damage grades based on the "scenario damage" simulations using the models for the total number of buildings in Le Teil are given in Figure 5. The simulations use different exposure, fragility, and site models, in addition to different ground motion modelling. The ground motion modelling labelled "ESHM20 GMF" in Figure 5, consists of ground motion fields for the sites of the exposure centroids generated with scenario analyses with the GMPE "KothaEtAl2020Site", as in Sec. 3.1, where the "Ritz et al." parameters (Table 3-1) are used for the rupture model. For the "SM GMF" model, we generate the ground motion fields, which are subsequently entered as input in the analyses with the OpenQuake Engine. These ground motion fields are generated using parameters for lognormal distributions of the PGA and the spectral acceleration at 0.3, 0.6, and 1.0 s, which are calculated based on a ShakeMap analysis. Moreover, the sampling of the ground motion fields use correlation models for the spatial correlation and the correlation between spectral accelerations at different periods (Baker and Jayaram, 2008; Jayaram and Baker, 2009).

In Figure 5, we may see the effect of the different models for the $V_{S30}$, the ground motion intensity, and the exposure. Figure 5 includes the probabilities of the damage grades from 8 different sources. Two of the sources consist of probabilities based on expert judgement ("Exp. judg.-based"), and probabilities based on our conversion of the damage observations to damage grades ("Observation-based"). The other 6 sources of the results in Figure 5 are "scenario damage" simulations, whose labels consist of 3 parts, each of the parts corresponding to a model in the simulation. The effect of the $V_{S30}$ model may be seen by comparing the 1st ("brgm $V_{S30}$ - ESHM20 GMF - brgm exp.") and the 5th set of results ("ESHM $V_{S30}$ - ESHM20 GMF - brgm exp.") in Figure 5. Significant differences are observed with respect to the damage grades 1 and 5. The model

"ESHM20 $V_{S30}$" leads to a lower mean probability for damage grade 5, while resulting to a higher probability for damage grade 1. The probabilities for the damage grades 2-3 for these two models do not present significant differences. We attribute the differences to the fact that there are lower $V_{S30}$ values in the "brgm $V_{S30}$" model (Table 3-5), which entails higher site amplification, higher ground motion intensities, and therefore higher probabilities for the higher damage grades.

*Table 3-6 Probabilities of EMS-98 damage grades conditioned on the building colour tag according to expert judgement*

| tag | P(DG1|tag) | P(DG2|tag) | P(DG3|tag) | P(DG4|tag) | P(DG5|tag) |
|---|---|---|---|---|---|
| Green | 0.80 | 0.20 | 0 | 0 | 0 |
| Yellow | 0 | 0.40 | 0.60 | 0 | 0 |
| Red | 0 | 0 | 0.55 | 0.40 | 0.05 |

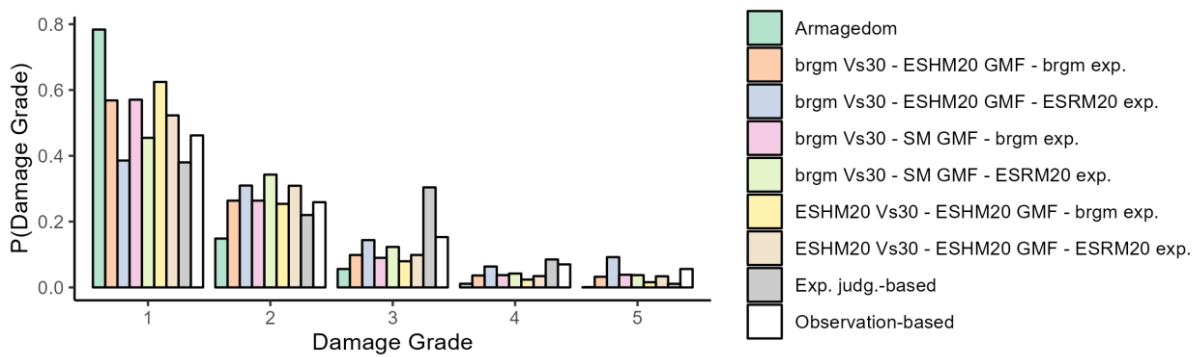

*Figure 5. Probabilities and number of buildings with EMS-98 damage grades for the configurations of the simulations with aggregated exposure including the total number of buildings in Le Teil*

## 4 Conclusion

Based on simulations of earthquake scenarios and ShakeMap analyses, we made comparisons of ground motion modelling, and comparisons of the estimated number of damages based on components of the ESRM20. The conversion of the samples of ground motion intensity measures, which were generated from the scenario simulations, with the FM2010 model led to estimations closer to the estimation by Schlupp et al. (2022). An explanation for the fact, that the AS2000 model led to macroseismic intensities lower than those based on the FM2010 model and the observation-based estimation, may be offered if the AS2000 model were created based on a data related to buildings, which are of different type and less vulnerable from those in Le Teil and from those in the data used for the development of the FM2010 model. In our opinion, this is a plausible explanation. Our estimation the furthest from this observation resulted from the conversion of the simulated Sa(1.0s) samples using the AS2000 model. We assume that Sa(1.0s), as a seismic ground motion intensity measure, is less related to the probability of damage in buildings, i.e., less "efficient" as defined by Luco and Cornell (2007), such as those in the municipality of Le Teil, which are mostly low to mid-rise and should have a first mode of vibration much lower than 1.0 s.

As far as the comparison with respect to the probabilities of the estimated damage is concerned, it highlighted the effect of the exposure model. The fact that the "brgm exp." model led to higher probabilities for the damage grades 3-5 than the ESHM20 exposure may be explained by the different percentage of masonry buildings in the two models; 47 % in the "brgm exp." Model versus 71 % in the "ESRM20 exp" model.

The proposed procedure based on the observed damages could be improved by introducing a probabilistic rule for the conversion of damage observations on the three-level colour tag (red, yellow, green) scale to the EMS-98 damage scale. Moreover, the fact that there is a need for this conversion leads us to recommend to future post-seismic surveys to record damage observations on the EMS-98 scale instead or in addition to the typical 3-level scale. Furthermore, our comparison with respect to the estimated damage highlighted the importance of the estimation of the probabilities of the damage grades in the buildings not included in the post-seismic survey. Such estimations could be made by relying on means such as satellite imaging, and rapid damage assessments based on low-cost sensors.

*Table A1 Selected ESRM20 fragility classes based on the building types in Le Teil according to the ESRM20*

| Original ESRM20 type | N. buildings | Selected ESRM20 frag. class | Class |
|---|---|---|---|
| CR+PC/LWAL+CDN/HBET:3-5 | 53 | CR_LDUAL-DUL_H4 | 1 |
| CR/LDUAL+CDL+LFC:4.0/HBET:3-5 | 7 | CR_LDUAL-DUL_H5 | 1 |
| CR/LDUAL+CDM+LFC:4.0/HBET:3-5 | 3 | CR_LDUAL-DUL_H6 | 1 |
| CR/LDUAL+CDL+LFC:4.0/HBET:6- | 3 | CR_LDUAL-DUL_H7 | 1 |
| CR/LDUAL+CDN/HBET:6- | 2 | CR_LDUAL-DUL_H8 | 1 |
| CR+PC/LWAL+CDN/HBET:6- | 1 | CR_LDUAL-DUL_H9 | 1 |
| CR/LDUAL+CDM+LFC:4.0/HBET:6- | 1 | CR_LDUAL-DUL_H10 | 1 |
| CR/LFINF+CDL+LFC:4.0/H:1 | 76 | CR_LFINF-CDL-10_H2 | 2 |
| CR/LFINF+CDL+LFC:4.0/H:2 | 67 | CR_LFINF-CDL-10_H2 | 2 |
| CR/LFINF+CDM+LFC:4.0/H:1 | 42 | CR_LFINF-CDM-10_H2 | 3 |
| CR/LFINF+CDM+LFC:4.0/H:2 | 37 | CR_LFINF-CDM-10_H2 | 3 |
| CR/LFINF+CDN/HBET:3-5 | 38 | CR_LFINF-CDL-15_H4 | 4 |
| CR/LFLS+CDN/HBET:6- | 9 | CR_LFINF-CDL-15_H4 | 4 |
| MUR+CL/LWAL+CDN/H:2 | 378 | MUR-CL99_LWAL-DNO_H2 | 5 |
| MUR+ST/LWAL+CDN/H:2 | 130 | MUR-CL99_LWAL-DNO_H2 | 5 |
| MUR+CL/LWAL+CDN/H:1 | 690 | MUR-CL99_LWAL-DNO_H1 | 6 |
| W/LWAL+CDN/H:1 | 100 | W_LFM-DUL_H2 | 7 |
| W/LWAL+CDN/H:2 | 43 | W_LFM-DUL_H2 | 7 |

*Table A2 Summary of the exposure based on the European Exposure model for the municipality of Le Teil*

| # | Selected ESRM20 class | N. of buildings |
|---|---|---|
| 1 | CR_LDUAL-DUL_H4 | 70 |
| 2 | CR_LFINF-CDL-10_H2 | 143 |
| 3 | CR_LFINF-CDM-10_H2 | 78 |
| 4 | CR_LFINF-CDL-15_H4 | 46 |
| 5 | MUR-CL99_LWAL-DNO_H2 | 508 |
| 6 | MUR-CL99_LWAL-DNO_H1 | 690 |
| 7 | W_LFM-DUL_H2 | 143 |

*Table A3 Summary of the BRGM/CCR exposure model for the municipality of Le Teil*

| # | Selected ESRM20 class | Number of buildings |
|---|---|---|
| 1 | CR_LFINF-CDL-10_H1 | 296 |
| 2 | CR_LFINF-CDL-10_H2 | 138 |
| 3 | CR_LFINF-CDL-15_H2 | 348 |
| 4 | CR_LFINF-CDL-15_H3 | 631 |
| 5 | CR_LFINF-CDL-15_H4 | 12 |
| 6 | CR_LFINF-CDM-0_H1 | 27 |
| 7 | CR_LFINF-CDM-10_H1 | 8 |
| 8 | MCF_LWAL-DUL_H2 | 127 |
| 9 | MCF_LWAL-DUL_H3 | 278 |
| 10 | MUR-STDRE_LWAL-DNO_H1 | 130 |
| 11 | MUR-STDRE_LWAL-DNO_H2 | 483 |
| 12 | MUR-STDRE_LWAL-DNO_H3 | 300 |